\documentclass[12pt,leqno]{article}
\usepackage{amsmath,amsthm}

\def\init{\setcounter{equation}{0}}
\newtheorem{theorem}{Theorem}[section]

\newcommand{\R}{{\bf R}}

\newtheorem{pro}{Proposition}[section]

\newcommand{\e}{{\varepsilon}}

\title{Perturbations of the Kerr black hole in the class of axisymmetric
artificial black holes.
\author{G.Eskin, \ \ \  Department of Mathematics, UCLA,\\ Los Angeles,
CA 90095-1555, USA. \ E-mail: eskin@math.ucla.edu}
}

\begin{document}

\maketitle

\begin{abstract}
Artificial black holes (called also acoustic or optical black holes)  are the black holes for 
the linear wave equation describing the wave propagation in a moving medium.  They attracted a considerable
interest of physicists who study them to better understand the black holes in general relativity.
We consider the case of stationary axisymmetric metrics and we show that the Kerr black
hole is not stable under perturbations in the class of all axisymmetric metrics. We describe families
 of axisymmetric metrics having black holes that are the perturbations of the Kerr black hole.  
We also show that the ergosphere can be determined by boundary measurements
and we prove the uniform boundness of the solution in the exterior  of the black hole when the event
horizon coincides with the ergosphere. 
\end{abstract}

\section{Introduction.}
\label{section 1}
\init

Consider the wave equation:

\begin{equation}                                               \label{eq:1.1}
\sum_{j,k=0}^n\frac{1}{\sqrt{(-1)^ng(x)}}\frac{\partial}{\partial x_j}\left(\sqrt{(-1)^ng(x)}g^{jk}(x)
\frac{\partial u(x_0,x)}{\partial x_k}\right)=0
\end{equation}
in $\Omega\times\R$,  where $\Omega$  is a smooth bounded domain in $\R^n, \ x_0\in \R$  is the 
time variable, $g(x)=\det[g_{jk}(x)]_{j,k=0}^n,\ [g_{jk}(x)]_{j,k=0}^n=   ([g^{jk}(x)]_{j,k=0}^n)^{-1}$
is a pseudo-Riemanian metric  with the Lorentz signature $(1,-1,...,-1)$.
We assume that $g^{jk}(x)$ are independent of $x_0$ and smooth in $\overline{\Omega}$.  We also assume that
\begin{equation}                                               \label{eq:1.2}
g^{00}(x)>0,\ \ \forall x\in\overline{\Omega}.
\end{equation}
Equation (\ref{eq:1.1})  describes the wave propagation in a moving medium.
We shall give two examples:

1)
The equation of propagation of light in a moving dielectric medium (the Gordon
equation,  cf. [G], [LP]).  In this case the metric has the form:
\begin{equation}                                   \label{eq:1.3}
g_{jk}(x)=\eta_{jk}+(n^{-2}(x)-1)v_jv_k,\ \ 0\leq j,k\leq 3,
\end{equation}
where $\eta_{jk}$  is the Lorenz metric tensor,  $\eta_{00}=1,\ \eta_{jj}=-1,\ 1\leq j\leq 3, 
\\ \eta_{jk}=0,\ j\neq k,\ n(x)=\sqrt{\e(x)\mu(x)}$  is  the refraction index,
$v_0(x)=\left(1-\frac{|w(x)|^2}{c^2}\right)^{-\frac{1}{2}},\\ 
v_j(x)=-\left(1-\frac{|w(x)|^2}{c^2}\right)^{-\frac{1}{2}}\frac{w_j}{c},\ 1\leq j\leq 3$,
is the four-velocity,  $w(x)=(w_1(x),w_2(x),w_3(x))$  is the velocity of the flow,  $c$
is the speed of light in the vacuum.

2)  The second example is the equation of acoustic waves in a moving fluid  
(cf. [V1]).  Here the metric tensor $[g_{jk}]_{j,k=0}^3$  has the form:

\begin{eqnarray}                                     \label{eq:1.4}
g_{00}=\frac{\rho}{c}(c^2-v^2),\ \ g_{0j}=g_{j0}=\frac{\rho}{c}v^j,\ 1\leq j\leq 3,
\\
\nonumber
g_{jk}=-\frac{\rho}{c}\delta_{jk},\ \ 1\leq j,k\leq 3,\ v^2=\sum_{j=1}^3(v^j)^2,
\end{eqnarray}
where $\rho$  is the density,  $c$  is the sound speed, $v=(v^1,v^2,v^3)$  is the velocity of 
the flow.
\qed

Let $S_0=\{x:S_0(x)=0\}$  be a closed smooth surface in $\R^n$.  The domain $\Omega_{int}\times\R$
where $\Omega_{int}$  is the interior of $S_0$  is called a black hole for (\ref{eq:1.1}) if
no signal from $\Omega_{int}\times\R$  can reach the exterior of $S_0\times\R$.
Analogously $\Omega_{int}\times\R$  is a white hole if no signal from the exterior
of $S_0\times\R$ can reach $\Omega_{int}\times\R$.  The 
surface $S_0\times\R$  is called the event horizon.  It is easy to show (see for example [E2]) 
that the interior of $S_0\times\R$ is a black hole if $S_0$ is a characteristic surface of
(\ref{eq:1.1}),  i.e.
\begin{equation}                                    \label{eq:1.5}
\sum_{j,k=1}^ng^{jk}(x)S_{0x_j}(x)S_{0x_k}(x)=0\ \ \ \mbox{when}\ \ S_0(x)=0
\end{equation}
and
\begin{equation}                                    \label{eq:1.6}
\sum_{j=1}^ng^{0j}(x)S_{0x_j}(x)<0,\ \ \ S_0(x)=0.
\end{equation}
Analogously $S_0\times \R$  is the boundary of a white hole if (\ref{eq:1.5}) holds and
\begin{equation}                                    \label{eq:1.7}
\sum_{j=1}^ng^{0j}(x)S_{0x_j}(x)>0,\ \ \ S_0(x)=0.
\end{equation}
In (\ref{eq:1.6}), (\ref{eq:1.7})  $S_{0x}$  is the outward normal to $S_0$.

Black (white)  holes for (\ref{eq:1.1}) are called artificial black (white) holes to
distinguish 
them
from the black (white) holes in the general relativity.  Physicists  
are interested in studying artificial black (white) holes to better understand the gravitational
black (white)  holes (cf. [NVV], [V1], [U]). 

The surface
\begin{equation}                                   \label{eq:1.8}   
g_{00}(x)=0
\end{equation}
is called the ergosphere.  We assume that (\ref{eq:1.8})
is a smooth closed surface,  $g_{00}(x)>0$  in the exterior of (\ref{eq:1.8}) and 
$g_{00}(x)<0$ in the interior of (\ref{eq:1.8}) near  $g_{00}(x)=0$.
An equivalent form of the equation (\ref{eq:1.8}) is
\begin{equation}                                 \label{eq:1.9}
\Delta(x)\stackrel{def}{=}\det [g^{jk}(x)]_{j,k=1}^n=0.
\end{equation}
If $\Delta=0$ is a characteristic surface,  i.e.
\begin{equation}                                 \label{eq:1.10}
\sum_{j,k=1}^ng^{jk}(x)\frac{\partial\Delta}{\partial x_j}\frac{\partial\Delta}{\partial x_k}=0,\ \ 
\Delta(x)=0,
\end{equation}
then  $\Delta\times\R$  is an event horizon and the interior 
of $\Delta\times\R$ is a black or white hole depending on the sign of 
$\sum_{j=1}^ng^{0j}(x)\Delta_{x_j}(x)$  when $\Delta(x)=0$  (cf. (\ref{eq:1.6}),  (\ref{eq:1.7})).
Note that any black or white hole is contained in 
the
closure of the interior of $\Delta\times\R$  
since 
$\Delta(x)>0$  outside of $\Delta\times\R$  and  any direction $(0,\xi),\ \xi\in\R^n\setminus\{0\}$
is not characteristic when $\Delta(x)>0$.  An example of a black hole whose event horizon 
is also 
the ergosphere is the Schwarzschild black hole where the metric
in Cartesian coordinates
 is (cf. [V2]):
\begin{eqnarray}                                 \label{eq:1.11}
\ \ \ \ \ \ \ \ ds^2=(1-\frac{2m}{R})dt^2-dx^2-dy^2-dz^2-\frac{4m}{R}dtdR-\frac{2m}{R}(dR)^2,
\\
\nonumber
 R^2=x^2+y^2+z^2.
\end{eqnarray}
It follows from (\ref{eq:1.11}) that $R=2m$  is the ergosphere and also the event horizon of a black                    
hole.  

In the next section we study the stability and nonstability of black holes in the case of
axisymmetric metrics.   The selebrated example of such metric is the Kerr metric [K].  In \S3  we show
that the boundary measurements allow to determine the ergosphere for any wave equation (\ref{eq:1.1}).
In the last section 4  we prove the uniform boundness of the solution of (\ref{eq:1.1}) in
the exterior of a black hole when the event horizon and the ergosphere are the same.   We shell call 
such black holes the Schwarzschild type black holes.  The proofs in \S 4 use substantially
 the ideas of the paper [DR].

\section{Axisymmetric metrics.}
\label{section 2}
\init

We shall use cylindrical coordinates $(\rho,z,\varphi)$ in $\R^3$  where
$$
\rho^2=x^2+y^2,\ \ x=\rho\cos\varphi,\ y=\rho\sin\varphi,\ z=z.
$$
A stationary axisymmetric metric in $\R^3\times\R$  is the metric that does not 
depend on $\varphi$ and $t$,  i.e. 
\begin{equation}                                     \label{eq:2.1}
ds^2=\sum_{j,k=0}^3g_{jk}(\rho,z)dy_jdy_k,
\end{equation}
where $g_{jk}(\rho,z)$ are smooth and even in $\rho$.  We use here the following notation:
 $y_0=t,\ y_1=\rho,\ y_2=z,
y_3=\varphi$.  Denote
$$
[g^{jk}(\rho,z)]_{j,k=0}^3=([g_{jk}(\rho,z)]_{j,k=0}^3)^{-1}.
$$
The ergosphere for (\ref{eq:2.1})  is given  by the equation
\begin{equation}                                       \label{eq:2.2}
g_{00}(\rho,z)=0,
\end{equation}
or, in equivalent form:
\begin{equation}                                       \label{eq:2.3}
\Delta(\rho,z)\stackrel{def}{=} \det[g^{jk}(\rho,z)]_{j,k=1}^3=0.
\end{equation}
The well known example of  axisymmetric metric is the Kerr metric
(cf. [K], [C]). 
The Kerr metric tensor in Kerr-Schild coordinates has the following form (see [V2]):
\begin{eqnarray}                                          \label{eq:2.4}
ds^2=dt^2-dx^2-dy^2-dz^2 \ \ \ \ \ \ \ \ \ \ \ \ \ \ \ \ \
\\
\nonumber
-\ \ \frac{2mr^3}{r^4+a^2z^2}
\left[dt+\frac{r(xdx+ydy)}{r^2+a^2}+
\frac{a(ydx-xdy)}{r^2+a^2}
+\frac{z}{r}dz\right]^2,
\end{eqnarray}
where
\begin{equation}                               \label{eq:2.5}
r(x,y,z)=\sqrt{\frac{(R^2-a^2)^2+\sqrt{(R^2-a^2)^2+4a^2z^2}}{2}},\ \ R^2=x^2+y^2+z^2.
\end{equation}
Therefore the equation of the ergosphere is
\begin{equation}                               \label{eq:2.6}
r^4+a^2z^2-2mr^3=0.
\end{equation}
It can be shown (see [V2])  that (\ref{eq:2.6}) consists of two curves
\begin{equation}                                \label{eq:2.7}
r-\left(m+\sqrt{m^2-\frac{a^2z^2}{r^2}}\right)=0,
\end{equation}
and
\begin{equation}                                \label{eq:2.8}
r-\left(m-\sqrt{m^2-\frac{a^2z^2}{r^2}}\right)=0,
\end{equation}
Equation (\ref{eq:2.7})  defines the outer ergosphere and (\ref{eq:2.8})  defines the inner 
ergosphere.  Note that (\ref{eq:2.8})  is not smooth  at $z=0,\ \rho^2=x^2+y^2=a^2$.

For general axisymmtric metric introduce
\begin{equation}                                 \label{eq:2.9}
\Delta_1(\rho,z)=\det[g^{jk}(\rho,z)]_{j,k=1}^2=g^{11}(\rho,z)g^{22}(\rho,z)-(g^{12}(\rho,z))^2.
\end{equation}
  We shall call $\Delta_1(\rho,z)=0$  the restricted ergosphere.

In the case of the Kerr metric the inverse of the metric tensor has the following
form in Cartesian coordinates
(see [V2]):
\begin{eqnarray}                              \label{eq:2.10}
\eta^{jk}+\frac{2mr^3}{r^4+a^2z^2}l^jl^k,
\\
\nonumber
(l^0,l^1,l^2,l^3)=\left(-1,\frac{rx+ay}{r^2+a^2},\frac{ry-ax}{r^2+a^2},\frac{z}{r}\right).
\end{eqnarray}
Therefore in $(\rho,z,\varphi)$  coordinates we have
\begin{equation}                              \label{eq:2.11}
g^{jk}(\rho,z)=\xi^{jk}+\frac{2mr^3}{r^4+a^2z^2}m^jm^k,
\end{equation}
where  $(m^0,m^1,m^2,m^3)=\left(-1,\frac{r\rho}{r^2+a^2},\frac{z}{r},\frac{-a}{r^2+a^2}\right),\
\xi^{jk}$  is $\eta^{jk}$  in the cylindrical coordinates,  i.e. 
$\xi^{jk}=0$  for  $j\neq k,\ \xi^{00}=1,\ \xi^{11}=\xi^{22}=-1,\ \xi^{33}=-\frac{1}{\rho^2},\ 
0\leq j,k\leq 3$.  In particular
\begin{eqnarray}                                   
\nonumber
g^{11}(\rho,z)=-1+\frac{2mr^5\rho^2}{(r^4+a^2z^2)(r^2+a^2)^2},
\\
\nonumber
g^{22}(\rho,z)=-1+
\frac{2mrz^2}{r^4+a^2z^2},\ \ \ 
g^{12}(\rho,z)=
\frac{2mr^3z\rho}{(r^4+a^2z^2)(r^2+a^2)}.
\end{eqnarray}
Therefore
\begin{equation}                                  \label{eq:2.12}
\Delta_1(\rho,z)=1-\frac{2mr^5\rho^2}{(r^4+a^2z^2)(r^2+a^2)^2}
-\frac{2mrz^2}{r^4+a^2z^2}.
\end{equation}
The equation $\Delta_1=0$  for the Kerr metric can be substantially simplified.

\begin{pro}                                       \label{prop:2.1}
The equation $\Delta_1(\rho,z)=0$  is equivalent 
to two equations $r-r_+=0$ and $r-r_-=0$  where 
$r_\pm=m\pm\sqrt{m^2-a^2}$.
\end{pro}
The proof of Proposition 2.1 is a straightforward computation.

It follows from (\ref{eq:2.5})  that 
equations $r-r_\pm=0$  has the following form in $(\rho,z)$  coordinates
\begin{equation}                                    \label{eq:2.13}
\frac{r_\pm}{2m}\rho^2+z^2=r_\pm^2.
\end{equation}
Multiplying $\Delta_1$  by $r^4+a^2z^2$ and expanding by the Taylor formula at $r=r_\pm$  we get
\begin{equation}                                    \label{eq:2.14}
r_\pm^4+a^2z^2-\frac{2mr_\pm^5\rho^2}{(2mr_\pm)^2}
-2mr_\pm z^2=O(r-r_\pm).
\end{equation}
Note that $r_\pm^2+a^2=2mr_\pm$.  Therefore (\ref{eq:2.14})  takes the form:
\begin{equation}                             \label{eq:2.15}
r_\pm^2-z^2-\frac{r_\pm\rho^2}{2m}=O(r-r_\pm).
\end{equation}
Compairing (\ref{eq:2.13})  and (\ref{eq:2.15})  we get that
$$
\Delta_1(\rho,z)=C_\pm(\rho,z)(r-r_\pm),
$$
where $C_\pm\neq 0$.
\qed

Note that $r-r_+=0$  and $r-r_-=0$ are the equations of the outer and the inner event horizons  
for the Kerr metric.  Therefore $\Delta_1(\rho,z)=0$  relates more explicitly  the outer
 and inner event horizons  to the metric.

In the case of the Kerr metric the points $\rho=0,z=\pm a$  are the only common to the 
ergospheres $r-\left(m\pm\sqrt{m^2-\frac{a^2z^2}{r^2}}\right)=0$  and to the event horizons
$r-r_\pm=0$.  All other points of $r-r_+=0$  are inside
$r-\left(m+\sqrt{m^2-\frac{a^2z^2}{r^2}}\right)=0$  and all other points of $r-r_-=0$  are outside of
$r-\left(m-\sqrt{m^2-\frac{a^2z^2}{r^2}}\right)=0$.

The same situation holds for any axisymmetric metric.  We consider first the ergosphere given
by a smooth closed curve $\Delta(\rho,z)=0$  and such that $\Delta>0$  outside of $\Delta=0$  
and $\Delta<0$  inside $\Delta=0$  near $\Delta=0$.  This situation   corresponds to the outer ergosphere 
for the Kerr metric.

\begin{pro}                                 \label{prop:2.2}  
Let  $\Delta=0$  be the ergosphere and 
$\Delta_1=0$  be the same 
as in (\ref{eq:2.9}),  $\Delta_1=0$ is a smooth closed curve and $\Delta_1>0$ outside of $\Delta_1=0$,
$\Delta_1<0$  inside $\Delta_1=0$ in a neighborhood of $\Delta_1=0$.  Then  any point $(\rho,z)$
on $\Delta_1=0$ either belongs to $\Delta=0$  or is inside $\Delta=0$.
\end{pro}

{\bf Proof:}  Let $\Delta(\rho,z)=0$.  Since the quadratic form 
$\sum_{j,k=1}^3g^{jk}(\rho,z)\eta_j\eta_k$  has one zero eigenvalue and two positive
when $\Delta=0$ there exists a smooth vector $b(\rho,z)=(b_1,b_2,b_3)$  such that
$$
\sum_{k=1}^3g^{jk}(\rho,z)b_k=0,\ \ 1\leq j\leq 3,\ \Delta(\rho,z)=0.
$$
Suppose $b_3(\rho,z)\neq 0$.  Then we shall show that the quadratic form
\begin{equation}                                  \label{eq:2.16}
\sum_{j,k=1}^2g^{jk}(\rho,z)\eta_j\eta_k
\end{equation}
is positive definite at $(\rho,z)$,  i.e. $\Delta_1(\rho,z)>0$.  Suppose the quadratic form
(\ref{eq:2.16})
has one negative eigenvalue.
Then there exists $(\eta_1,\eta_2)$
such that $\sum_{j,k=1}^2g^{jk}(\rho,z)\eta_j\eta_k<0$.
Take $\eta_3=0$.  Then 
$\sum_{j,k=1}^3g^{jk}(\rho,z)\eta_j\eta_k=\sum_{j,k=1}^2g^{jk}(\rho,z)\eta_j\eta_k<0$
and this contradicts the fact that $[g^{jk}(\rho,z)]_{j,k=1}^3$
has no negative eigenvalues.  Suppose the quadratic form (\ref{eq:2.16}) has one zero eigenvalue.
Then there exists $(\eta_1^{(1)},\eta_2^{(1)})$
such that
$\sum_{j,k=1}^2g^{jk}(\rho,z)\eta_j^{(1)}\eta_k^{(1)}=0$.  Put $\eta_3^{(1)}=0$.
Then
$\sum_{j,k=1}^3g^{jk}(\rho,z)\eta_j^{(1)}\eta_k^{(1)}=0$,   i.e. $(\eta_1^{(1)},\eta_2^{(2)},0)$
is also a null-vector of the matrix $[g^{jk}(\rho,z)]_{j,k=1}^3$.
This is a contradiction since the null-space of the matrix $[g^{jk}(\rho,z)]_{j,k=1}^3$  is
one-dimensional and vectors $(b_1(\rho,z),b_2(\rho,z),b_3(\rho,z))$  and
$(\eta_1^{(1)},\eta_2^{(1)},0)$  are linearly independent.  We proved that the 
quadratic form (\ref{eq:2.16}) is positive definite at $(\rho,z)$ and therefore $\Delta_1(\rho,z)>0$.
This means that $(\rho,z)$ is outside of the curve $\Delta_1=0$.  When $b_3(\rho,z)=0$  
then $(\rho,z)$ is a common point of
$\Delta=0$  and $\Delta_1=0$.
\qed

{\bf Remark 2.1}  In the case of the Kerr metric there is also the inner ergosphere 
$\Delta^-=r-\left(m-\sqrt{m^2-\frac{a^2z^2}{r^2}}\right)=0$.
In this case in a neighborhood of  $\Delta^-=0$  we have that $\Delta^->0$  inside of $\Delta^-=0$
and $\Delta^-<0$  outside of $\Delta^-=0$.  The same proof as in 
Proposition \ref{prop:2.2} gives that $\Delta_1^-=0$  is outside of the inner ergosphere 
where $\Delta_1^-=0$ is the restricted ergosphere (cf. (\ref{eq:2.9}).
\qed

Suppose $\psi_1=0$ is a 
closed
characteristic curve,  i.e. $\{\psi_1=0\}\times S^1\times\R$ is
an event horizon.  
We say 
that the   black or the white hole with the event horizon $\{\psi_1=0\}\times S^1\times\R$ is stable 
if any smooth family of axysymmetric  metrics $g_\e=[g_{\e jk}]_{j,k=0}^3,0\leq \e\leq\e_0$ such that 
$g_{jk0}=g_{jk}$  has a smooth family of event horizons $\{\psi_1^\e=0\}\times S^1\times\R$  where 
$\psi_1^\e=\psi_1$  when $\e=0$.  Otherwise we say that $\{\psi_1=0\}\times S^1\times\R$  is
unstable.

Let $\Delta_1=0$  be the restricted ergosphere (see (\ref{eq:2.9})) We shall prove that if
$\Delta_1=0$  is also a characteristic curve then $\{\Delta_1=0\}\times S^1\times\R$  is unstable event
horison.  Consider, for the simplicity, the axisymmetric metrics of the form 
\begin{equation}                                \label{eq:2.17}
g^{jk}(\rho,z)=\xi^{jk}+v^j(\rho,z)v^k(\rho,z),\ \ \ 0\leq j,k\leq 3.
\end{equation}
Note that the Kerr metric (cf. (\ref{eq:2.11})) and also the metrics (\ref{eq:1.3}),\ (\ref{eq:1.4})  have the form 
(\ref{eq:2.17}).

\begin{pro}                                  \label{prop:2.3}
Consider a metric of the form (\ref{eq:2.17}).  Let $\Delta_1=0$
be the same as in (\ref{eq:2.9})
and let $\Delta_1=0$  be a characteristic curve.  Then the event horizon $\{\Delta_1=0\}\times S^1\times\R$  is
unstable when we perturb $[g^{jk}]$  in the class of all axisymmetric metrics 
$g_\e^{jk}=\xi^{jk}+v_\e^jv_\e^k$  of the form (\ref{eq:2.17}).
\end{pro}

{\bf Proof:}  Let $U_0$  be a small neighborhood of some point 
$(\rho_0,z_0)\in \{\Delta_1=0\}$.  In $U_0$  we can represent $v^1$  and $v^2$ 
in the form:  $v^1=\beta(\rho,z)\cos\alpha(\rho,z),\ \ v^2=\beta(\rho,z)\sin\alpha(\rho,z)$,
where $\alpha,\beta$  are smooth, $\beta>0$.  For the perturbed metric we also have:
\begin{equation}                             \label{eq:2.18}
v_\e^1=\beta_\e(\rho,z)\cos\alpha_\e(\rho,z),\ \ 
v_\e^2=\beta_\e(\rho,z)\sin\alpha_\e(\rho,z),\ \ \ 0\leq \e\leq\e_0,
\end{equation}
where
$\alpha_\e,\beta_\e >0$  are arbitrary smooth functions in $U_0$  satisfying the
conditions $\alpha_0(\rho,z)=\alpha(\rho,z),\ \ \beta_0(\rho,z)=\beta(\rho,z)$.
Since 
$$
[g^{jk}]_{j,k=1}^2=
\left[\begin{array}{ll}-1+(v^1)^2\ \ \ \ &v^1v^2\\ \ \ \ \ 
v^1v^2  & -1+(v^2)^2
\end{array}
\right]
$$
we have that the restricted ergosphere $\Delta_1^\e=0$  for $g_\e$  has the form 
$\Delta_1^\e=1-(v_\e^1)^2-(v_\e^2)^2=0,$  i.e.
\begin{equation}                                \label{eq:2.19}
\Delta_1^\e=1-\beta_\e^2(\rho,z)=0.
\end{equation}
If $\Delta_1^\e=0$  is a characteristic curve then
\begin{equation}                                \label{eq:2.20}
\sum_{j,k=1}^2g_\e^{jk}\Delta_{1j}^\e\Delta_{1k}^\e=0\ \ \mbox{on}\ \ \Delta_1^\e=0.
\end{equation}
Here $\Delta_{11}^\e=\Delta_{1\rho}^\e,\ \Delta_{12}^\e=\Delta_{1z}^\e$.  Since  the quadratic 
form (\ref{eq:2.20})  has one zero and one positive eigenvalue,  the equation
(\ref{eq:2.20})  is equivalent to
\begin{eqnarray}                                \label{eq:2.21}
(-1+(v_\e^1)^2)\Delta_{1\rho}^\e+v_\e^1v_\e^2\Delta_{1z}^\e=0,
\\
\nonumber
v_\e^1v_\e^2\Delta_{1\rho}^\e+(-1+(v_\e^2)^2)\Delta_{1z}^\e=0,
\end{eqnarray}
when $\Delta_1^\e=0$.

Substituting (\ref{eq:2.18})  in (\ref{eq:2.21})  we get when (\ref{eq:2.19})
holds:
\begin{equation}                                    \label{eq:2.22}
\beta_{\e z}\cos\alpha_\e-\beta_{\e \rho}\sin\alpha_\e=0.
\end{equation} 
Since $\alpha_\e,\beta_\e$  are arbitrary  when $\e>0$  we can choose $\alpha_\e,\beta_e$  
such that (\ref{eq:2.22})
is different from zero when $\beta_\e=1$  for all $\e>0,(\rho,z)\in U_0$ and
$\alpha_\e=\alpha_0,\ \beta_\e=\beta_0$  for all $0\leq \e\leq \e_0$  when
$(\rho,z)\notin U_0$.  Therefore  $\Delta_1^\e=\Delta_1$  outside of $U_0$.  However 
$\Delta_1^\e=0$ 
  is not a closed characteristic curve when $0<\e\leq\e_0$.

We shall show now that $g_\e$  has no event horizon near $\Delta_1=0$.  
Denote by $\Delta_{1\delta}^\e=0$
a curve inside $\Delta_1^\e=0$  such that the distance between  $\Delta_1^\e=0$
and $\Delta_{1\delta}^\e=0$  is equal to $\delta$.
Suppose there exists $\psi_1^\e=0$  between $\Delta_1^\e$  and  $\Delta_{2\delta}^\e=0$ 
such that $\psi_1^\e=0$  is a closed characteristic curve for $0<\e\leq \e_0$  and
$\psi_1^0=\Delta_1$.
Consider the characteristic curves inside $\Delta_1^\e=0$  starting at $\Delta_1^\e=0$.  They are
solutions of the differential equations (4.9) in [E2].  Outside of $U_0$,  i.e. when $\Delta_1^\e=
\Delta_1=0$,  
these solutions of (4.9) in [E2]  are
parabola-like   and    they cross $\Delta_{1\delta}^\e=0$,
where $\delta>0$  is small and independent of $\e$.  There is no other characteristic curves 
outside of $U_0$  between $\Delta_1^\e=0$  and $\Delta_{1\delta}^\e=0$.
Therefore $\psi_1^\e=0$  must coincide  with $\Delta _1^\e=0$  outside of $U_0$.  Note that 
$\psi_1^\e=0$  and  $\Delta_1^\e=0$  differ in $U_0$  since the normal to $\Delta_1^\e=0$  is
not characteristic at any point of $U_0$.

Denote by $Q_\e$  the domain between $\Delta_1^\e=0$  and  $\psi_1^\e=0$.
Let $P_1$  and $P_2$  be the points of intersection of $\Delta_1^\e=0$  and $\psi_1^\e=0$.
Assign to the characteristic curve $\psi_1^\e=0$  the direction from $P_1$  to $P_2$.
Consider characteristic curves in $Q_\e$  belonging to the same family as $\psi_1^\e=0$
and starting on $\Delta_1^\e=0$  close to $P_1$.  They will be close to $\psi_1^\e=0$  and will
end on $\Delta_1^\e=0$  close to $P_2$.  This will be a contradiction  because the
normals to $\Delta_1^\e=0$  are not characteristic in $U_0$  and therefore  all characteristic 
curves between $P_1$  and $P_2$  belonging to the same family are eithee start on $\Delta_1^\e=0$  
or end on $\Delta_1^\e=0$.  Therefore there is no closed characteristic curve near $\Delta_{1\e}=0$,
i.e.  there is no event horizon for $g_\e$  in the vicinity of
$\{\Delta_1=0\}\times S^1\times\R$.
\qed

Proposition \ref{prop:2.3}  implies the instability of the outer and inner horizons
for the Kerr metric.

In the following proposition we shall show that there exists a rich class of axisymmetric 
deformations $[g_{\e jk}]$  of the Kerr metric that has a smooth family of event horizons 
$\{\psi_1^\e=0\}\times S^1\times\R,\ 0\leq\e\leq\e_0$,  such that $\psi_1^0=\psi_1$.

\begin{pro}                                  \label{prop:2.4}
Let $\{\psi_1=0\}\times S^1\times\R$  be the Kerr outer event horizon.  
Let $\psi_1^\e(\rho,z)=0,\ 0\leq \e\leq \e_0,$  be an arbitrary
family of closed, even in $\rho$, smooth curves such that $\psi_1^0=\psi_1$.  
Then there exists a smooth family $g_\e$  of axisymmetric metrics of the
from (\ref{eq:2.17})  such that $\{\psi_1^\e=0\}\times S^1\times\R$  is the
event horizon corresponding to 
$g_\e,0\leq \e\leq\e_0,\ \ g_0=g$  is the Kerr metric and $\psi_1=\psi_1^0$.
\end{pro}

{\bf Proof:}  Denote
by $a_\e(\rho,z)$  the solution of the following eiconal equation  near $\psi_1=0$:
\begin{equation}                                 \label{eq:2.23}
\left(\frac{\partial a_\e}{\partial\rho}\right)^2+
\left(\frac{\partial a_\e}{\partial z}\right)^2=a_\e(\rho,z)
\end{equation}
such that 
\begin{equation}                            \label{eq:2.24}
a_\e=1 \ \ \mbox{on}\ \      \psi_1^\e=0
\end{equation}
and
\begin{equation}                             \label{eq:2.25}
\left(\frac{\partial a_\e}{\partial\rho}
\frac{\partial \psi_1^\e}{\partial \rho}+
\frac{\partial a_\e}{\partial z}
\frac{\partial \psi_1^\e}{\partial z}
\right)
>0 \ \ \mbox{on} \ \ \psi_1^\e=0.
\end{equation}

Note that
$\left(\frac{\partial a_\e}{\partial\rho},
\frac{\partial a_\e}{\partial z}\right) $ and
$\left(\frac{\partial \psi_1^\e}{\partial\rho},
\frac{\partial \psi_1^\e}{\partial z}\right)$
are colinear on $\psi_1^\e=0$  since they both are normals to  $\psi_1^\e=0$.
The condition (\ref{eq:2.25}) implies 
that they have the same direction and $a_\e>1$  when $\psi_1^\e>0,\ a_\e<1$  when 
$\psi_1^\e<0$.
The solution of (\ref{eq:2.23}), (\ref{eq:2.24}), (\ref{eq:2.25}) exists in some neighborhood
of                                                                                                                                             
$\psi_1=0$.  Since 
$\psi_1^0=\psi_1=0$  is characteristic curve   we have (cf. (\ref{eq:2.17})):
\begin{equation}                                      \label{eq:2.26}
 \left(\frac{\partial a_0}{\partial\rho}\right)^2+
\left(\frac{\partial a_0}{\partial z}\right)^2=
\left(v^1(\rho,z)\frac{\partial a_0}{\partial\rho}+
v^2(\rho,z)\frac{\partial a_0}{\partial z}\right)^2,\ \ \psi_1^0=0.
\end{equation}
Note that for the Kerr metric we have (cf. (\ref{eq:2.11})):
\begin{equation}                                   \label{eq:2.27}
v^1=\sqrt{\frac{2mr}{r^4+a^2z^2}}\ \frac{r\rho}{r^2+a^2},\ \ \ v^2=\sqrt{\frac{2mr}{z^4+a^2z^2}}
\ \frac{z}{r}.
\end{equation}  
Since $\psi_1^0=0$  is also an ergosphere we have
\begin{equation}                           \label{eq:2.28}
(v^1)^2+(v^2)^2=1\ \ \ \mbox{on}\ \ \psi_1^0=0,
\end{equation}
since $\Delta_1=(1-(v^1)^2)(1-(v^2)^2)-(v^1v^2)^2=1-(v^1)^2-(v^2)^2$.
Therefore (\ref{eq:2.26}), (\ref{eq:2.28})  imply that
$$
\frac{\partial a_0}{\partial \rho}=v^1, \ \ \frac{\partial a_0}{\partial z}=v^2\ \ \ \
\mbox{on}\ \ \psi_1=0.
$$
Construct smooth vectors
$v_\e^1(\rho,z),v_\e^2(\rho,z)$  in a neighborhood of $\psi_1=0$
such that $v_0^1=v^1,\ v_0^2=v$
and $v_\e^1=\frac{\partial a_\e}{\partial\rho},\ v_\e^2=\frac{\partial a_\e}{\partial z}$
on $\psi_1^\e=0$.
After finding $v_\e^1,v_\e^2$   in a neighborhood of $\psi_1=0$  we
extend them for all $(\rho,z)$  and choose also   components 
$v_\e^0(\rho,z),v_\e^3(\rho,z)$
of the tensor $g_\e^{ij}$
to get a metric that coincides with the Kerr metric for $\e=0$  near $\psi_1=0$.
Then $\psi_1^\e=0$  is the restricted ergosphere and a characteristic curve.  Therefore 
$\{\psi_1^\e=0\}\times S^1\times\R$  is a family of event horizons   for the metrics $g_\e$.

{\bf Remark 2.2}
It follows from the proof of Proposition \ref{prop:2.4} that for any metric of the form 
(\ref{eq:2.17}) there exists a family of metric $g_\e$  with a family of event horizons  
 smoothly dependent on $\e,\ 0\leq \e\leq\e_0$.  Note that $\psi_1^\e$ are chosen
arbitrary with the only restriction that $\psi_1^\e=0$  is a smooth 
closed curve even in $\rho$  and $\psi_1^0=\psi_1$.

{\bf Remark 2.3}
When we consider the perturbation $g_\e$  of the metric $g_0$ we are getting a family of
perturbed ergospheres  $\psi^\e=0$.  If  $\psi^0=\psi=0$  was a smooth, even in $\rho$  curve in
$(\rho,z)$  plane then  $\psi^\e=0$  will be also smooth curves  for small $\e>0$.
This is the case of the outer ergosphere for the Kerr metric.  The inner ergosphere
$r-(m-\sqrt{m^2-\frac{a^2z^2}{r^2}})=0$  is not smooth  at $\rho^2=a^2,  z=0$.
Therefore the perturbed ergosphere may be also not smooth and have points of self-intersection.
However the perturbation of the inner event horizon $\psi_1^-=r-(m-\sqrt{m^2-a^2})=0$  
will proceed as in outer event horizon case
since
$r-(m-\sqrt{m^2-a^2})=0$  is  smooth.
In particular  as in Proposition \ref{prop:2.3}  we have that $r-(m-\sqrt{m^2-a^2})=0$  is unstable.

Consider 
the changes in the proof of Proposition \ref{prop:2.4}.   
 We chose $a_\e$  as a solution of (\ref{eq:2.23})
near $r=r_-,\ a_\e=1$  on $r=r_-$  and $a_\e>1$  outside of $r=r_-,\ a_\e<1$ 
inside of $r=r_-$.  The rest of arguments in the proof of Proposition \ref{prop:2.4}
is the same 
and we get a family $\psi_1^\e=0,\psi_1^0=r-(m-\sqrt{m^2-a^2})=0$  such 
that $\{\psi_1^\e=0\}\times S^1\times\R$
are the event hozizons for a family of metrics $g_\e$ of the form (\ref{eq:2.17}) such that
$g_0$  coincides with the Kerr metric near $r-(m-\sqrt{m^2-a^2})=0.$

In the following Proposition we consider the case of not necessarily \\ axisymmetric metrics.
\begin{pro}                                    \label{prop:2.5}
Let $S(x)=0$  be arbitrary smooth closed surface in $\R^n$.  There exists a stationary metric 
$([g^{jk}(x)]_{j,k=0}^n)^{-1}$  such that $S(x)=0$ is an ergosphere and $\{S(x)=0\}\times\R$  is
an event horizon.
\end{pro}

{\bf Proof:}  Let, as before, $a(x)$  be the solution of
$$
\sum_{j=1}^n\left(\frac{\partial a}{\partial x_n}\right)^2=a
$$
such that $a=1$  on $S(x)=0$,  $a<1$ outside  of $S(x)=0$  and $a>1$  inside $S(x)=0$
near $S(x)=0$.
Construct $v^j(x),1\leq j\leq n,$  such that $v^j(x)=\frac{\partial a}{\partial x_j}$ on
$S(x)=0, \ \sum_{j=1}^n(v^j)^2<1$  outside of $S(x)=0$  and $\sum_{j=1}^n(v^j)^2>1$
inside $S(x)=0$  near $S(x)=0$.  Consider the following inverse to the metric tensor 
(cf. (\ref{eq:2.17}))
$$
g^{jk}(x)=\eta^{jk}-v^j(x)v^k(x)\ \ \mbox{for}\ \ 1\leq j,k\leq n,
$$
$$
g^{j0}(x)=v^j(x),\ \ g^{00}=1.
$$
  We have 
$$
\sum_{k=1}^n\left(\frac{\partial a}{\partial x_k}\right)^2
=\left(\sum_{k=1}^n\frac{\partial a}{\partial x_k}v^k\right)^2\ \ \mbox{on}\ \ S(x)=0
$$
since $v^k=\frac{\partial a}{\partial x_k}$  on $S(x)=0$.  Therefore  $\ \ S(x)=0$  
is a characteristic surface
since $\left(\frac{\partial a}{\partial x_1},...,\frac{\partial a}{\partial x_j}\right)$
is a normal to $S(x)=0$.
Since $\sum_{j=1}^n(v^j)^2=1$  on $S(x)=0$  the quadratic form
$\sum_{j=1}^n\xi_j^2-\left(\sum_{j=1}^nv^j\xi_j\right)^2\geq 0$  on $S(x)=0$  and
$(\xi_1,...,\xi_n)=(v^1...,v^n)$   
is the only null-vector.  Therefore $S(x)=0$  is an ergosphere.

\section{Determination of the ergosphere by the boundary measurements.}
\label{section 3}
\init
Let $u(x_0,x)$ be  the solution of (\ref{eq:1.1})  in a cylinder $\Omega\times \R$ satisfying the
zero initial conditions
\begin{equation}                                   \label{eq:3.1}
u=0\ \ \ \mbox{for} \ \ x_0 \ll 0,\ x\in\Omega,
\end{equation}
and the boundary condition
\begin{equation}                                   \label{eq:3.2}
u{\LARGE |}_{\partial\Omega\times\R}=f.
\end{equation}
We assume  that $\Omega$  is a smooth bounded  domain in $\R^n$,  $f$  is 
a smooth function with a compact support in
$\partial\Omega\times\R$.  The solution of the initial boundary value problem 
(\ref{eq:1.1}),
(\ref{eq:3.1}), (\ref{eq:3.2})
exists and is unique assuming that $\partial\Omega$  is not characteristic and (\ref{eq:1.2}) holds.
Denote by $\Lambda f$  the following operator (the Dirichlet-to-Neumann operator):
\begin{equation}                                    \label{eq:3.3}
\Lambda f=\sum_{j,k=1}^ng^{jk}(x)\frac{\partial u}{\partial x_j}\nu_k(x)\left.\left(\sum_{p,r=1}^n
g^{pr}(x)\nu_p\nu_r\right)^{-\frac{1}{2}}\right|_{\partial\Omega\times\R},
\end{equation}
where
$(\nu_1(x),...,\nu_n(x))$
is the unit outward normal to $\partial\Omega$.  Let $\Gamma$  be any open subset of $\partial\Omega$.
We say that the boundary measurements on $\Gamma\times(0,T)$ are taken if we know $\Lambda f$
on $\Gamma\times(0,T)$  for any $f$  with support in $\overline{\Gamma}\times[0,T]$.  The inverse
boundary value problem is the recovery of $[g^{jk}(x)]_{j,k=0}^n$ and the rest of $\partial\Omega$,
i.e. $\partial\Omega\setminus\Gamma$,  knowing the boundary measurements  on $\Gamma\times(0,T)$
for some $\Gamma$  and $T>0$.

Let  
\begin{equation}                                   \label{eq:3.4}
\hat{x}=\varphi(x),\ \ \hat{x}_0=x_0+a(x),
\end{equation}
be the change of variables such that $\hat x=\varphi(x)$  is a diffeomophism of 
$\overline{\Omega}$  onto a new domain $\overline{\hat \Omega},\ \ a(x)\in C^\infty(\overline{\Omega})$.
We assume that 
\begin{equation}                                    \label{eq:3.5}
\varphi(x)=x  \ \ \ \mbox{on}\ \ \ \overline{\Gamma} \ \ \mbox{and}\ \ \
a(x)=0\ \ \mbox{on}\ \ \overline{\Gamma}.
\end{equation}
Note that (\ref{eq:3.4})  transforms (\ref{eq:1.1})  to an equation of the same form with
the new  tensor $[\hat g^{jk}(\hat x)]$ isometric to the old one.
It follows  from (\ref{eq:3.5})   that
the DN operator $\Lambda$  does not change under the change of variable (\ref{eq:3.4}),
(\ref{eq:3.5}).  Therefore one can hope only to recover the metric up to
the change of variables (\ref{eq:3.4}),  (\ref{eq:3.5}).

If there exists an event horizon inside $\Omega\times \R$  then 
we can not determine the metric inside the event horizon since any change of metric inside
will not change  the boundary measurements.
  But we can try to recover the
event horizon itself (up to diffeomorphism (\ref{eq:3.4}), (\ref{eq:3.5})).  This problem is
still open.  We can only prove that the boundary measurements allow to determine the ergosphere. 
\begin{theorem}                                      \label{theo:3.1}
Consider the wave equation (\ref{eq:1.1}).
  Assume (\ref{eq:1.2})  holds and $\partial\Omega$  is 
not characteristic.
Let the ergosphere $\Delta(x)=\det[g^{jk}(x)]_{j,k=1}^n=0$  be a smooth closed surface, $\Delta(x)>0$
in $\overline \Omega$  outside $\Delta(x)=0$.  Let $\Gamma$  be an open subset of $\partial\Omega$.
Then the boundary measurements on $\Gamma\times(0,+\infty)$ determine $\Delta(x)=0$  up
to the change of variables (\ref{eq:3.4}),  (\ref{eq:3.5}).
\end{theorem}
Note that it does not matter whether the ergosphere is an event horizon or not.

{\bf Proof:}  The proof of Theorem 2.3 in [E1] can be applied to prove Theorem \ref{theo:3.1}.
As in [E1]  we start with the determination of the metric in a small neighborhood of $\Gamma$  
and gradually continue to recover the metric deeper  in $\Omega$.  As we proceed 
the time interval $(0,T)$  needed to reach a point $x\in \Omega$ increases as the point approaches 
the ergosphere.
Let $x=x(\sigma),\ 0\leq \sigma\leq 1,$  be an arbitrary curve in $\Omega\cap\Omega_{ext}$,  
where $\Omega_{ext}$  is the exterior of $\Delta(x)=0,\ x(0)\in\Gamma,\ x(1)\in
\{\Delta=0\}$.   Consider the restriction of the metric to the two-dimensional surface 
$H=\{x=x(\sigma),\sigma\in[0,1]\}\times \R.$
We get 
$$
ds^2=a_{00}(\sigma)dx_0^2+2a_{01}(\sigma) dx_0d\sigma + a_{11}(\sigma)d\sigma^2
$$
$$
=a_{11}(\sigma)(d\sigma-\lambda_+(\sigma)dx_0)(d\sigma-\lambda_-(\sigma)dx_0),
$$
where
$a_{11}(\sigma)\neq 0$ on $[0,1]$.  Since (\ref{eq:1.1})  is hyperbolic,  $g_{00}(x)>0$  in 
$\Omega\setminus\overline{\Omega}_{ext}$  and $g_{00}(x)=0$  on $\Delta(x)=0$,
we have that $\lambda_+(\sigma)>0,\lambda_-(\sigma)<0$  on $[0,1)$  and either
$\lambda_+(\sigma)$  or $\lambda_-(\sigma)$  is equal
to zero at $\sigma=1$.

Fix $\sigma_0\in(0,1)$  and consider curves $x_0=x_0^+(\sigma),x_0=x_0^-(\sigma),\sigma\in[0,\sigma_0]$
where $\frac{dx_0^-}{d\sigma}=\lambda_+(\sigma),\ x_0^-(\sigma_0)=0,\ 
\frac{dx_0^+}{d\sigma}=\lambda_-(\sigma),\ x_0^+(\sigma_0)=0$.  The triangle $T$  bounded by $x_0=
x_0^+(\sigma),\ x_0=x_0^-(\sigma),\ 0\leq \sigma\leq\sigma_0$,
and the vertical line $l=\{x=x(0),x_0^-(0)\leq x_0\leq x_0^+(0)\}$  is the domain of the unique
continuation for $l$.  When $\sigma_0\rightarrow 1$  the length 
of $l$, i.e. $x_0^+(0)-x_0^-(0)$  tends to $+\infty$.  
 These arguments help to explain why
the unique determination of the 
surface $\Delta(x)=0$  requires the measurements on the infinite interval $\Gamma\times(0,+\infty)$.
Similar arguments play important role in the proof
of Theorem 2.3 in [E1] (cf. also  [ER]). 

\section{Uniform boundness of solutions  in the exterior of the Schwartzschild type black hole.}
\label{section 4}
\init

Let $S_0=\{x:S_0(x)=0\}$
be a closed smooth surface in $\R^n$  such that $S_0(x)=0$  is an ergosphere and an event horizon,
i.e.  $S_0(x)=0$  is also a characteristic surface (cf. \S 1).     
Let $\nu(x)=(\nu_1(x),...,\nu_n(x))$ be the outward unit normal to $S_0(x)=0$.
We assume that (cf. (1.6))
\begin{equation}                              \label{eq:4.1}
\sum_{j=1}^n g^{0j}(x)\nu_j(x)<0\ \ \ \mbox{on}\ \ \  S_0,
\end{equation}
i.e. $\Omega_{int}\times\R$  is a black hole.  We denote the equation (\ref{eq:1.1}) by
$\qed_g u=0$  and consider the initial value problem
\begin{equation}                               \label{eq:4.2}
\qed_g u=0 \ \ \ \mbox{in}\ \ \Omega_{ext}\times[0,+\infty),
\end{equation}
\begin{equation}                               \label{eq:4.3}
u(0,x)=u_0(x),\ \  u_{x_0}(0,x)=u_1(x),\ \ x\in\Omega_{ext}.
\end{equation}
We assume that
$u_0\in H^2(\Omega_{ext}),\ u_1\in H^1(\Omega_{ext})$  and
$u_0(x),u_1(x)$  decay when $|x|\rightarrow\infty$.  Assume also that  
$g^{jk}(x)=\delta_{jk}+O(\frac{1}{|x|})$.
\begin{theorem}                                   \label{theo:4.1}
Under the above assumptions the solution $u(x_0,x)$  of (\ref{eq:4.2}),  (\ref{eq:4.3})  is 
uniformly bounded
\begin{equation}                                  \label{eq:4.4}
|u(x_0,x)|\leq C \ \ \ \mbox{for all}\ \ x\in \overline{\Omega}_{ext}\subset\R^3,\ x_0\geq 0.
\end{equation}
\end{theorem}

{\bf Proof:}
Let $(u,v)=\int_{\Omega_{ext}}u(x)v(x)\sqrt{|g(x)|}dx$.
Integrating  the integral   \\
 $0=\int_0^T(\qed_gu,u_{x_0})dx_0$ 
by parts in $x_j,1\leq j\leq n$,  
we get
\begin{eqnarray}                                          \label{eq:4.5}
\ \ \ \ \ \ \ 
E_T(u)-E_0(u)=\int_0^T\int_{S_0}\left(\sum_{j=1}^ng^{j0}(x)\nu_j(x)\right)
u_{x_0}^2(x_0,x)\sqrt{g}dsdx_0
\\
\nonumber
+\int_0^T\int_{S_0}\sum_{j,k=1}^ng^{jk}(x)\nu_j(x)u_{x_k}u_{x_0}\sqrt{|g|}dsdx_0,
\end{eqnarray}
where 
\begin{equation}                                       \label{eq:4.6}
E_{x_0}(u)=\frac{1}{2}\int_{\Omega_{ext}}\left[g^{00}u_{x_0}^2(x_0,x)
-\sum_{j,k=1}^ng^{jk}(x)u_{x_j}u_{x_k}\right]\sqrt{|g|}dx,
\end{equation}
and $ds$  is the Euclidean surface element,  $g^{00}>0$.
We used in (\ref{eq:4.5})  that
$$
(\ \sum_{j=1}^n\frac{1}{\sqrt{|g|}}
\frac{\partial}{\partial x_j}\left(\sqrt{|g|}g^{j0}\frac{\partial u}{\partial x_0}\right)
+
\sum_{j=1}^n\frac{1}{\sqrt{|g|}}
\frac{\partial}{\partial x_0}\left(\sqrt{|g|}g^{j0}\frac{\partial u}{\partial x_j}\right),u_{x_0})
$$
$$
=-\int_{S_0}\left(\sum_{j=1}^ng^{j0}\nu_j\right)u_{x_0}^2\sqrt{|g|}ds.
$$
Note that $\nu(x)$  is an outward normal to $\Omega_{int}$
and therefore it is an inward normal to $\Omega_{ext}$.
The quadratic form $-\sum_{j,k=1}^ng^{jk}(x)\xi_j\xi_k$  has one zero  eigenvalue and $(n-1)$  positive
eigenvalues on $S_0$.  Therefore (cf. \S 2):
\begin{equation}                                  \label{eq:4.7}
\sum_{j=1}^ng^{jk}(x)\nu_j(x)=0,\ \ 1\leq k\leq n,\ \ \mbox{on}\ \  S_0,
\end{equation}
and the last integral in (\ref{eq:4.5}) is equal to zero.  Therefore
\begin{equation}                                   \label{eq:4.8}
E_T(u)-\int_0^T\left(\sum_{j=1}^ng^{0j}\nu_j\right)u_{x_0}^2\sqrt{|g|}dsd_{x_0}=E_0(u).
\end{equation}
It follows from (\ref{eq:4.1})  that
\begin{equation}                                     \label{eq:4.9}
E_T(u)\leq E_0(u),\ \ \ 
-\int_0^T\left(\sum_{j=1}^ng^{0j}\nu_j\right)u_{x_0}^2\sqrt{|g|}dsd_{x_0}\leq E_0(u).
\end{equation}
Since 
$$
-\sum_{j,k=1}^ng^{jk}(x)u_{x_j}u_{x_k}\geq 0 \ \ \ 
\mbox{and}\ \ -\sum_{j,k=1}^ng^{jk}(x)u_{x_j}u_{x_k}\geq C_\delta\sum_{j=1}^n u_{x_j}^2
$$
when $d(x,S_0)\geq \delta$  we  have
\begin{equation}                                      \label{eq:4.10}
\int_{\Omega_\delta^{(1)}}\sum_{j=0}^n u_{x_j}^2\sqrt{|g|}dx\leq C_\delta E_{x_0}(u)\leq C_\delta E_0(u),\ \ 
\forall x_0\geq 0,
\end{equation}
where  $\Omega_\delta=\Omega_{ext}\cap\{x:d(x,S_0)<\delta\},\ \ 
\Omega_\delta^{(1)}=\Omega_{ext}\setminus\Omega_\delta,\ d(x,S_0)$  is the distance from $x$  to $S_0$.

Since the coefficients
 of $\qed_g$  are independent of $x_0$  we can differenciate  $\qed_g u=0$  in $x_0$  to obtain
\begin{equation}                                 \label{eq:4.11}
E_{x_0}(u_{x_0^m})\leq E_0(u_{x_0^m}),\ \ \forall m\geq 1.
\end{equation}
Let $\chi_0(x)\in C^\infty(\Omega_{ext}),\ \chi_0(x)=0$  when $d(x,S_0)\leq \delta,\ \chi_0=1$
when $d(x,S_0)\geq 2\delta$.  We have
$$
0\ =\ \chi_0(x)\qed_g u\ =\ -L\chi_0 u-f,
$$
where 
$$
L=-\sum_{j,k=1}^n\frac{\partial}{\partial x_j}g^{jk}\frac{\partial}{\partial x_k},\ \ \ \ \
f=-\chi_0g^{00}u_{x_0^2}+L_0u_{x_0}+L_1u,
$$
$L_k$  are differential operators in $\frac{\partial}{\partial x}$  of order 1,  $k=0,1$.  
By the elliptic regularity we  have 
$$\|\chi_0(x)u(x_0,x)\|_2\leq C\|f\|_0\leq C(\ \|u_{x_0^2}\|_0+\|u_{x_0}\|_{1,\Omega_\delta^{(1)}}+
\|u\|_{1,\Omega_\delta^{(1)}}\ ).
$$
Using (\ref{eq:4.10}), (\ref{eq:4.11})  we get
\begin{equation}                                 \label{eq:4.12}
\|\chi_0(x)u\|_2\leq C(E_0(u)+E_0(u_{x_0})).
\end{equation}
Now we shall study $\qed_gu=0$  near $S_0(x)=0$.
Let $x^{(0)}\in S_0$  and  let $U_0$  be a small neighborhood
of $x^{(0)}$.  Assume for the definiteness that $\frac{\partial S_0}{\partial x_n}\neq 0$
in $U_0$  and make a change of variables
\begin{equation}                                         \label{eq:4.13}
\hat x_n=S(x),\ \ \hat x_k=x_k,\ \ 0\leq k\leq n-1.
\end{equation}
Let 
$$
\qed_{\hat g}\hat u=\sum_{j,k=0}^n\frac{1}{\sqrt{|\hat g|}}
\frac{\partial}{\partial \hat x_j}\left(\sqrt{|\hat g|}\hat g^{jk}(\hat x)
\frac{\partial\hat u}{\partial \hat x_k}\right)=0
$$
be the operator  $\qed_g$ in new coordinates,\  $u(x_0,x)=\hat u(x_o,\hat x)$.  Note that
\begin{equation}                                     \label{eq:4.14}
\hat g^{nn}(\hat x)=\sum_{j,k=1}^n g^{jk}(x)S_{0x_j}(x)S_{0x_k}(x),
\end{equation}
i.e. $\hat g^{nn}=0$  when $\hat x_n=0$.  Also  we have
\begin{eqnarray}                                 \label{eq:4.15}
\hat g^{nk}=\sum_{j=1}^ng^{jk}(x)S_{0x_j},\ \ \ 0\leq k\leq n-1,
\\
\nonumber
\hat g^{j0}=g^{j0},\ \ 0\leq j\leq n-1,\ \ \ \hat g^{jk}= g^{jk},\ \ 1\leq j,k\leq n-1. 
\end{eqnarray}
It follows from (\ref{eq:4.1}) and (\ref{eq:4.7})  that
\begin{equation}                                 \label{eq:4.16}
\hat g^{n0}<0,\ \ \hat g^{nk}=0,\ \ 1\leq k\leq n-1,\ \ \ \mbox{when}\ \ \hat x_n=0.
\end{equation}
Note also that
\begin{equation}                                 \label{eq:4.17}
-\hat g^{nn}\geq C\hat x_n,\ \ \ -\hat g_{\hat x_n}^{nn}\geq C>0\ \ \mbox{in}\ 
\ U_0\cap\overline{\Omega}_{ext}.
\end{equation}
The quadratic form $-\sum_{j,k=1}^ng^{jk}u_{x_j}u_{x_k}$  has the following 
lower bound
in $(x_0,\hat x)$  coordinates in $U_0$:
\begin{equation}                                    \label{eq:4.18}
-\sum_{j,k=1}^n\hat g^{jk}(\hat x)\hat u_{\hat x_j}\hat u_{\hat x_k}\geq 
C\left( \hat x_n\left(\frac{\partial\hat u}{\partial \hat x_n}\right)^2
+\sum_{j=1}^{n-1}\left(\frac{\partial \hat u}{\partial x_j}\right)^2\right).
\end{equation}
\qed

Now we shall describe another classical identity different from (\ref{eq:4.5}).

Denote 
$$
Hu=\sum_{j=1}^ng^{j0}u_{x_j}.
$$
We shall study $$ 0=\int_0^T(\qed_gu,g^{00}u_{x_0}+Hu)dx_0$$.
We have
\begin{eqnarray}                                   \label{eq:4.19}
\int_0^T(\ \sum_{j=0}^n\frac{1}{\sqrt{|g|}}
\frac{\partial}{\partial x_0}\left(\sqrt{|g|}g^{j0}\frac{\partial u}{\partial x_j}\right)
+\sum_{j=1}^n\frac{1}{\sqrt{|g|}}
\frac{\partial}{\partial x_j}\left(\sqrt{|g|}g^{j0}\frac{\partial u}{\partial x_0}\right),g^{00}u_{x_0}+Hu)dx_0
\nonumber
\\
\ \ \ \ \ \ \ \ \ \ \ \ \ \ \ \ 
\stackrel{def}{=} I_{11}+I_{12}+I_{13},
\ \ \ \ \ \ \ \ \ \ \ \ \ \ \ \ \ \ \ \ \ \ \ \ \ \ \ \ \ \ \ \ \ \ \ \ \ \ \ \
\ \ \ \ \ \ \ \ \ \ \ \ \ \ \ \ \ \ \ \ 
 \end{eqnarray}
where
\begin{equation}                                \label{eq:4.20}
I_{11}=\frac{1}{2}\int_{\Omega_{ext}}(g^{00}u_{x_0}+Hu)^2\sqrt{|g|}\ dx\big|_0^T,
\end{equation}
and we use the notation $a{\big|}_0^T=a(T)-a(0)$.

Furthermore
\begin{eqnarray}                                   \label{eq:4.21}
I_{12}=\int_0^T(\ \frac{1}{\sqrt{|g|}}
 \sum_{j=1}^n
\frac{\partial}{\partial x_j}\left(\sqrt{|g|}g^{j0}\frac{\partial u}{\partial x_0}\right),Hu)dx_0
\\
\nonumber
= \frac{1}{2}\int_{\Omega_{ext}}(Hu)^2\sqrt{|g|}\ dx{\big |}_0^T +
\int_0^T\int_{\Omega_{ext}}\left(\sum_{j=1}^n
\frac{\partial}{\partial x_j}(\sqrt{|g|}g^{j0})\right)u_{x_0}Hu\ dx\ dx_0.
\end{eqnarray}
Finally,
\begin{eqnarray}                                   \label{eq:4.22}
I_{13}=
\int_0^T(\ \frac{1}{\sqrt{|g|}}
 \sum_{j=1}^n
\frac{\partial}{\partial x_j}\left(\sqrt{|g|}g^{j0}\frac{\partial u}{\partial x_0}\right),
g^{00}u_{x_0})dx_0
\\
\nonumber
=
\frac{1}{2}
\int_0^T\int_{\Omega_{ext}}\sum_{j=1}^n
\sqrt{|g|}g^{j0}g^{00}\frac{\partial}{\partial x_j}u_{x_0}^2\ dx\ dx_0
\\
\nonumber
+
\int_0^T\int_{\Omega_{ext}}\sum_{j=1}^n
\left(\frac{\partial}{\partial x_j}(\sqrt{|g|}g^{j0})\right)g^{00}u_{x_0}^2\ dx\ dx_0
\\
\nonumber
=-\ \frac{1}{2}
\int_0^T\int_{\Omega_{ext}}\sum_{j=1}^n
\frac{\partial}{\partial x_j}(\sqrt{|g|}g^{j0}g^{00})u_{x_0}^2\ dx\ dx_0
\\
\nonumber
+
\int_0^T\int_{\Omega_{ext}}\left(\sum_{j=1}^n
\frac{\partial}{\partial x_j}(\sqrt{|g|}g^{j0})\right)g^{00}u_{x_0}^2\ dx\ dx_0
\\
\nonumber
-\ \frac{1}{2}
\int_0^T\int_{S_0}g^{00}\left(\sum_{j=1}^ng^{j0}\nu_j)\right)u_{x_0}^2\sqrt{|g|}\ ds\ dx_0
\end{eqnarray}
Consider now
\begin{equation}                                       \label{eq:4.23}
I_2=
\int_0^T(\ \frac{1}{\sqrt{|g|}}
 \sum_{j,k=1}^n
\frac{\partial}{\partial x_j}\left(\sqrt{|g|}g^{jk}\frac{\partial u}{\partial x_k}\right),
g^{00}u_{x_0})dx_0.
\end{equation}
Integrating by parts we get
\begin{eqnarray}
\nonumber
I_2=
-\int_0^T\int_{\Omega_{ext}}\sum_{j,k=1}^n\sqrt{|g|}g^{jk}u_{x_k}\frac{\partial}{\partial x_j}
(g^{00}u_{x_0})\ dx\ dx_0
\\
\nonumber
-\int_0^T\int_{S_0}\sum_{j,k=1}^ng^{jk}\nu_ju_{x_k}g^{00}u_{x_0}\sqrt{|g|}\ ds\ dx_0
\\
\nonumber
=
-\ \frac{1}{2}\int_{\Omega_{ext}}\sum_{j,k=1}^ng^{jk}u_{x_j}u_{x_k}g^{00}\sqrt{|g|}\ dx{\big|}_0^T
\\
\nonumber
-\int_0^T\int_{\Omega_{ext}}\sum_{j,k=1}^ng^{jk}u_{x_k}
\frac{\partial g^{00}}{\partial x_j} u_{x_0}\sqrt{|g|}\ dx\ dx_0
\\
\nonumber
-\int_0^T\int_{S_0}\sum_{j,k=1}^ng^{jk}\nu_ju_{x_k}g^{00}u_{x_0}\sqrt{|g|}\ ds\ d_0.
\end{eqnarray}

It remains to compute the integral
\begin{eqnarray}                                          \label{eq:4.24}
I_3=
\int_0^T(\ 
 \sum_{j,k=1}^n\frac{1}{\sqrt{|g|}}
\frac{\partial}{\partial x_j}\left(\sqrt{|g|}g^{jk}\frac{\partial u}{\partial x_k}\right),
Hu)dx_0
\\
\nonumber
=-\int_0^T\int_{\Omega_{ext}}\sum_{j,k=1}^n\sqrt{|g|}\ g^{jk}\frac{\partial u}{\partial x_k}
\frac{\partial}{\partial x_j} Hu\ dx\ dx_0
\\
\nonumber
-\int_0^T\int_{S_0}\sum_{j,k=1}^ng^{jk}u_{x_k}\nu_j Hu\sqrt{|g|}\ ds\ d_0.
\end{eqnarray}
Note that 
\begin{equation}                                            \label{eq:4.25}                                  
\frac{\partial}{\partial x_j} Hu=\sum_{p=1}^n\frac{\partial g^{p0}}{\partial x_j}u_{x_p}
+\sum_{p=1}^ng^{p0}u_{x_jx_p}.
\end{equation}
Integgrating by parts in  $x_p$  we get
\begin{eqnarray}                                    \label{eq:4.26}
-\int_{\Omega_{ext}}\sum_{j,k=1}^ng^{jk}u_{x_k}\frac{\partial}{\partial x_j} Hu\sqrt{|g|}\ dx
\\
\nonumber
= -\ \frac{1}{2}\int_{\Omega_{ext}}\sum_{p=1}^ng^{p0}\sum_{j,k=1}^ng^{jk}
\sqrt{|g|}\frac{\partial}{\partial x_p}(u_{x_j}u_{x_k})\ dx 
\\
\nonumber
-
\int_{\Omega_{ext}}\sum_{j,k=1}^ng^{jk}u_{x_k}\sum_{p=1}^ng_{x_j}^{p0}u_{x_p}\sqrt{|g|}\ dx
\\
\nonumber
=
\frac{1}{2}\int_{\Omega_{ext}}\sum_{p=1}^n\frac{\partial}{\partial x_p}
(g^{p0}\sqrt{|g|})\sum_{j,k=1}^ng^{jk}u_{x_j}u_{x_p}) dx
\\
\nonumber
+ \frac{1}{2}\int_{\Omega_{ext}}\sum_{j,k=1}^n(Hg^{jk})u_{x_j}u_{x_k}\sqrt{|g|}\ dx
\\
\nonumber
-\ \int_{\Omega_{ext}}\sum_{j,k=1}^ng^{jk}u_{x_k}
\left(\sum_{p=1}^n\frac{\partial g^{p0}}{\partial x_j^k}\right)\sqrt{|g|}\ dx
\\
\nonumber
+\frac{1}{2}\int_{S_0}\left(\sum_{p=1}^n g^{p0}\nu_p\right)\sum_{j,k=1}^ng^{jk}u_{x_j}u_{x_k}
\sqrt{|g|}\ ds.
\end{eqnarray}
Combining (\ref{eq:4.19})-(\ref{eq:4.26})  we get the following equality
\begin{eqnarray}                                       \label{eq:4.27}
0=\frac{1}{2}\int_{\Omega_{ext}}
[(g^{00}u_{x_0}+Hu)^2+(Hu)^2
\\
\nonumber
-  g^{00}\sum_{j,k=1}^ng^{jk}u_{x_j}u_{x_k}
]\sqrt{|g|}\ dx{\big|}_0^T
+B(u)+T(u),
\end{eqnarray}
where the boundary terms
$B(u)$  have the following form
\begin{eqnarray}                                 \label{eq:4.28}
B(u)= - \frac{1}{2}\int_0^T\int_{S_0}g^{00}\left(\sum_{j=1}^ng^{j0}\nu_j\right)u_{x_0}^2
\sqrt{|g|}\ ds\ dx_0
\\
\nonumber
\int_0^T\int_{S_0}
-\sum_{j,k=1}^ng^{jk}\nu_ju_{x_k}(g^{00}u_{x_0}+Hu)\sqrt{|g|}\ ds\ dx_0
\\
\nonumber
+\frac{1}{2}\int_0^T\int_{S_0}\left(\sum_{p=1}^ng^{p0}\nu_p\right)
\sum_{j,k=1}^ng^{jk}u_{x_j}u_{x_k}\sqrt{|g|}\ ds\ dx_0,
\end{eqnarray}
and the lower order terms $T(u)$  has the following form
\begin{eqnarray}                                        \label{eq:4.29}
T(u)=\int_0^T\int_{\Omega_{ext}}\left(\sum_{j=1}^n\frac{\partial}{\partial x_j}(\sqrt{|g|}
\ g^{j0})\right)
u_{x_0}Hu\ dx\ dx_0                   
\\
\nonumber
+\int_0^T\int_{\Omega_{ext}}\left[-\frac{1}{2}\sum_{j=1}^n
\frac{\partial}{\partial x_j}
(\sqrt{|g|}g^{j0}g^{00})
+\sum_{j=1}^n\frac{\partial}{\partial x_j}(\sqrt{|g|}g^{j0})g^{00}\right]
u_{x_0}^2\ dx\ dx_0
\\
\nonumber
-\int_0^T\int_{\Omega_{ext}}
\sum_{j,k=1}^ng^{jk}u_{x_k}\frac{\partial g^{00}}{\partial x_j}u_{x_0}\sqrt{|g|}\ dx\ dx_0
\\
\nonumber
+\ \frac{1}{2}\int_0^T\int_{\Omega{ext}}\sum_{p=1}^n\frac{\partial}{\partial x_p}(g^{p0}\sqrt{|g|})
\sum_{j,k=1}^ng^{jk}u_{x_j}u_{x_k}\ dx\ dx_0
\\
\nonumber
-\int_0^T\int_{\Omega_{ext}}\sum_{j,k=1}^ng^{jk}u_{x_k}
\left(\sum_{p=1}^n\frac{\partial g^{p0}}{\partial x_j}u_{x_p}\right)
\sqrt{|g|}\ dx\ dx_0
\\
\nonumber
+\ \frac{1}{2}\int_0^T\int_{\Omega_{ext}}\sum_{j,k=1}^n(Hg^{jk})u_{x_j}u_{x_k}\sqrt{|g|}\ dx\ dx_0.
\end{eqnarray}
It follows
from (\ref{eq:4.7})  that the second integral in (\ref{eq:4.28})
is equal to zero.  Furthermore,  $B(u)>0$  since $\sum_{j=1}^n g^{j0}\nu_j<0$  and 
$\sum_{j,k=1}^n g^{jk}u_{x_j}u_{x_k}\leq 0$.
Moreover,  (\ref{eq:4.11})  and (\ref{eq:4.18})  imply that 
\begin{equation}                                  \label{eq:4.30}
B\geq C\|u\|_{1,S_0\times(0,T)}^2,
\end{equation}
where                                  
$\|u\|_{1,S_0\times(0,T)}$ is the norm in $H^1(S_0\times(0,T))$.  Let
\begin{eqnarray}                                 \label{eq:4.31}
e_{x_0}(u)=-\sum_{j,k=1}^ng^{jk}u_{x_j}u_{x_k},\ \ \ e_{1x_0}(u)=(Hu)^2+e_{x_0}(u),
\\
\nonumber
E_1(x_0)=\frac{1}{2}\int_{\Omega_{ext}}[(g^{00}u_{x_0}+Hu)^2+(Hu)^2
-\sum_{j,k=1}^ng^{jk}u_{x_j}u_{x_k}]\ 
\sqrt{|g|}\ dx.
\end{eqnarray}
Note that 
\begin{equation}                                  \label{eq:4.32}
C_1(\|u\|_1^2+\|u_{x_0}\|_0^2)\leq E_1(x_0)\leq C(||u\|_1^2+\|u_{x_0}\|_0^2),
\end{equation}
where
$\|v\|_m$  is the norm in $H^m(\Omega_{ext})$.  Note also that the integral
\begin{equation}                                    \label{eq:4.33}
\int_{\Omega_{ext}} e_{1x_0}(u)\sqrt{|g|}\ dx
\end{equation}
is equivalent to $\|u\|_1^2$.
\qed

We shall need the estimate of $T(u)$  from below.  Denote the integrals in (\ref{eq:4.29})  by
$T_j(u),\ 1\leq j\leq 6$. 
We have 
$$
|T_2(u)|+|T_4(u)| \leq C\int_0^T\int_{\Omega_{ext}}\left(g^{00}u_{x_0}^2-
\sum_{j,k=1}^n g^{jk}u_{x_j}u_{x_k}\right)\sqrt{|g|}\ dx\ dx_0
$$
$$
=C\int_0^T E_{x_0}(u)dx_0\leq CTE_0(u).
$$
Therefore 
\begin{equation}                                        \label{eq:4.34}
T_2(u)+T_4(u)\geq -CTE_0.
\end{equation}

Now we shall  prove that
\begin{equation}                                        \label{eq:4.35}
T_6(u)\geq \int_0^T(dE_1(x_0)-DE_{x_0}(u))dx_0,
\end{equation}
where $d,D$  are some constants.
In local coordinates (\ref{eq:4.13})  the quadratic form
$\sum_{j,k=1}^n H(g^{jk})u_{x_j}u_{x_k}$  is equal to
\begin{equation}                                        \label{eq:4.36}
\sum_{j,k=1}^n\hat a^{jk}\hat u_{\hat x_j}\hat u_{\hat x_k},
\end{equation}
where 
\begin{equation}                                       \label{eq:4.37}
\hat a^{jk}=\hat H\hat g^{jk}
=\sum_{p=1}^n\hat g^{p0}\frac{\partial\hat g^{jk}}{\partial \hat x_p}.
\end{equation}
In particular,
$$
\hat a^{nn}=\hat g^{n0}\frac{\partial \hat g^{nn}}{\partial \hat x_n}
+\sum_{p=1}^{n-1}\hat g^{p0}\frac{\partial \hat g^{nn}}{\partial x_p}.
$$
Since $\hat g^{n0}<0$  and $\frac{\partial \hat g^{nn}}{\partial \hat x_n}<0,\ \hat g^{nn}=0$
when $\hat x_n=0$ (cf. (\ref{eq:4.16}), (\ref{eq:4.17}))  we have that
\begin{equation}                                     \label{eq:4.38}
\hat a^{nn}>0\ \ \ \mbox{near}\ \ \ \hat x_n=0.
\end{equation}
Therefore for $0\leq \hat x_n<\delta$  we obtain 
\begin{equation}                                      \label{eq:4.39}
\sum_{j,k=1}^n\hat a^{jk}\hat u_{\hat x_j}\hat u_{\hat x_k}\geq C_0\hat u_{\hat x_n}^2-
C_1\sum_{j=1}^{n-1}\hat u_{x_j}^2
\geq C_0\hat u_{x_n}^2-C_2\hat e_{x_0}(\hat u),
\end{equation}
where $\hat e_{x_0}(\hat u)$  is the quadratic form $e_{x_0}(u)$  in $\hat x$-coordinates.
To obtain (\ref{eq:4.39}) we used that
$$
\left|\sum_{j=1}^{n-1}\hat a^{nj}\hat u_{\hat x_n}\hat u_{\hat x_j}\right|
\leq \frac{\e}{2}\hat u_{\hat x_n}^2+\frac{1}{2\e}\left(\sum_{j=1}^{n-1}\hat a^{nj}\hat u_{\hat x_j}\right)^2.
$$     
Since $\hat u_{\hat x_n}^2+\hat e_{x_0}(\hat u)$  is a positive definite quadratic form for
$0\leq x_n\leq \delta$  it is equivalent to $\hat e_{1x_0}(\hat u)$.  Therefore
\begin{equation}                                 \label{eq:4.40}
\hat u_{\hat x_n}^2=\hat u_{x_n}^2+\hat e_{x_0}(\hat u)-\hat e_{x_0}(\hat u)\geq
C_3\hat e_{1x_0}(\hat u)-\hat e_{x_0}(\hat u).
\end{equation}
Combining (\ref{eq:4.39}) and (\ref{eq:4.40})  we obtain
\begin{equation}                                  \label{eq:4.41}
\sum_{j,k=1}^n(\hat H\hat g^{jk})\hat u_{\hat x_j}\hat u_{\hat x_k}
\geq C_0C_3\hat e_{1x_0}-(C_2+1)\hat e_{x_0}.
\end{equation}
Decreasing $C_3>0$ and increasing $C_2$  if needed we can make (\ref{eq:4.41})  hold in
local coordinates of a neighborhood of any point of $S_0$.  Therefore (\ref{eq:4.41})  holds in
the original coordinates $(x_1,...,x_n)$  in $\Omega_\delta$,  i.e.
\begin{equation}                                  \label{eq:4.42}
\sum_{j,k=1}^n(Hg^{jk})u_{x_j}u_{x_k}\geq C_0C_3 e_{1x_0}-(C_2+1)e_{x_0}.
\end{equation}
Since $e_{x_0}(u)$  is positive definite in $\Omega_{ext}\setminus\Omega_\delta$   we can,
by increasing $C_2$,  achieve that (\ref{eq:4.42}) holds in $\overline{\Omega}_{ext}$.
Note that $u_{x_0}^2+e_{1x_0}(u)$  and $(g^{00}u_{x_0}+Hu)^2+e_{1x_0}$  are positive definite 
quadratic forms
in $(u_{x_0},...,u_{x_n})$  and therefore they are equivalent.  In particular:
$$
u_{x_0}^2+e_{1x_0}\geq C_4[(g^{00}u_{x_0}+Hu)^2+e_{1x_0}].
$$
Therefore
\begin{eqnarray}                                     \label{eq:4.43}
\sum_{j,k=1}^n(Hg^{jk})u_{x_j}u_{x_k}\geq C_0C_3e_{1x_0}-(C_2+1)e_{x_0}
\\
\nonumber
=C_0C_3((u_{x_0}^2+e_{1x_0})-C_0C_3u_{x_0}^2-(C_2+1)e_{x_0}
\\
\nonumber
\geq C_0C_3C_4[(g^{00}u_{x_0}+Hu)^2+e_{1x_0}]-C_5(g^{00}u_{x_0}^2+e_{x_0}).
\end{eqnarray}
Integrating (\ref{eq:4.43})   over $\Omega_{ext}\times (0,T)$  we get (\ref{eq:4.35}).

To estimate $T_1(u)$  we use again that 
$$
|u_{x_0}Hu|\leq \frac{\e}{2}(Hu)^2+\frac{1}{2\e}u_{x_0}^2.
$$
Note that $(Hu)^2\leq e_{1x_0}(u)$.   Therefore
\begin{equation}                                  \label{eq:4.44}
T_1(u)\geq -\int_0^T C\e E_1(x_0)dx_0-\frac{C}{\e}\int_0^TE_{x_0}(u)dx_0.
\end{equation}
Since $-\sum_{j,k=1}^ng^{jk}\xi_j\xi_k\geq 0$  we can use the generalized Cauchy-Schwartz inequality
to estimate $T_5(u)$:
\begin{eqnarray}                                            \label{eq:4.45}
\ \ \ \ \ \ \ \ \ \ 
\left|\sum_{j,k=1}^n g^{jk}u_{x_k}l_j\right|\leq \left(-\sum_{j,k=1}^ng^{jk}u_{x_j}u_{x_k}\right)^{\frac{1}{2}}
\cdot\left(-\sum_{j,k=1}^ng^{jk}l_jl_k\right)^{\frac{1}{2}}
\\
\nonumber
\leq \frac{\e}{2}\left(-\sum_{j,k=1}^ng^{jk}l_jl_k\right)+ 
\frac{1}{2\e}\left(-\sum_{j,k=1}^n g^{jk}u_{x_j}u_{x_k}\right)
\\
\nonumber
\leq C\e \sum_{j=1}^nu_{x_j}^2+\frac{1}{2\e}\left(-\sum_{j,k=1}^ng^{jk}u_{x_j}u_{x_k}\right).
\end{eqnarray}
Here $l_j=\sum_{p=1}^n\frac{\partial g^{p0}}{\partial x_j}u_{x_p}$.  Applying (\ref{eq:4.45}) to
$T_5(u)$  we get:
\begin{equation}                                    \label{eq:4.46}
T_5(u)\geq -C\e\int_0^TE_1(x_0)dx_0-C_\e E_0T.
\end{equation}
The estimate of $T_3(u)$  is similar.
Collecting (\ref{eq:4.34}),  (\ref{eq:4.35}),  (\ref{eq:4.44}),  
(\ref{eq:4.46}) and replacing the time interval $(0,T)$  by $(\tau,t), \ 0\leq \tau\leq t\leq  T$,
we get from  (\ref{eq:4.27}) (cf. [DR]):
\begin{equation}                                      \label{eq:4.47}
E_1(t)-E_1(\tau)+B(u)+c\int_\tau^t E_1(x_0)dx_0-CE_0(t-\tau)\leq 0.
\end{equation}
We can drop $B$  in (\ref{eq:4.47}) since $B>0$.  Dividing by $t-\tau$  and taking the limit 
when $\tau\rightarrow t$  we get
\begin{equation}                                      \label{eq:4.48}
\frac{dE_1(t)}{dt}+cE_1(t)\leq CE_0.
\end{equation}
Solving (\ref{eq:4.48})  we obtain
\begin{equation}                                     \label{eq:4.49}
E_1(t)\leq E_1(0)+\frac{C}{c}E_0,\ \ \ \forall t\geq 0.
\end{equation}
Since $\qed_g$   commutes with $\frac{\partial}{\partial x_0}$  
the inequality (\ref{eq:4.49}) holds when $u(x_0,x)$ is replaced  by $u_{x_0^m}$
for any $m\geq 1$   (cf.  (\ref{eq:4.11})).
Now we shall use (\ref{eq:4.49}) to estimate $u(x_0,x)$  in $\Omega_\delta$.  We have,
in local coordinates:
\begin{eqnarray}                                     \label{eq:4.50}
0=\qed_{\hat g}\hat u =-\hat L\hat u + \hat L_1\hat u,
\\
\nonumber
\mbox{where}\ \ \ \
\hat L \hat u=-\sum_{j,k=1}^n\frac{\partial }{\partial \hat x_j}
\hat g^{jk}\frac{\partial \hat u}{\partial \hat x_k},\ \ \hat L_1=\qed_{\hat g}+\hat L.
\end{eqnarray}
Since $\hat u\in H^1(\Omega_\delta)$  and $\hat u_{x_0}\in H^1(\Omega_\delta)$  we have that
$L_1\hat u\in L^2(\Omega_\delta)$  and
\begin{equation}                                 \label{eq:4.51}
\|\hat L_1\hat u\|_0\leq C,
\end{equation}                            
where in (\ref{eq:4.51}) and below constants are independent of $x_0$  (cf. (\ref{eq:4.49})).

Let $(\hat u,\hat v)=\int_{\Omega_\delta}\hat u(x)\hat v(x)dx$  and let $\chi(x)\in C_0^\infty(U_0)$.
It follows from (\ref{eq:4.50}) that
\begin{equation}                              \label{eq:4.52}
0=(\chi\frac{\partial}{\partial x_p}\hat L\hat u,\chi\frac{\partial\hat u}{\partial x_p})
-(\chi\frac{\partial}{\partial x_p} L_1\hat u,\chi\frac{\partial\hat u}{\partial x_p}),\ \ \ 
1\leq p\leq n-1.
\end{equation}
Integrating by parts 
we have:
\begin{equation}                                  \label{eq:4.53}
|(\chi\frac{\partial}{\partial x_p} L_1\hat u,\chi\frac{\partial\hat u}{\partial x_p})|
=|(L_1\hat u,\frac{\partial}{\partial x_p} \chi^2\frac{\partial\hat u}{\partial x_p})|
\leq \e\|\frac{\partial^2}{\partial  x_p^2}(\chi\hat u)\|_0^2+C_\e.
\end{equation}
Commuting $\chi\frac{\partial}{\partial  x_p}$  and  $L$
and  using that (cf. (\ref{eq:4.18}))
\begin{equation}                                 \label{eq:4.54}
(\hat L\chi\frac{\partial\hat u}{\partial  x_p},\chi\frac{\partial\hat u}{\partial  x_p})
\geq
C\int_{\Omega_\delta}
\left(\hat x_n\left|\frac{\partial}{\partial \hat x_n}\chi\frac{\partial \hat u}{\partial  x_p}\right|^2+
\sum_{j=1}^{n-1}\left|\frac{\partial}{\partial  x_j}\chi\frac{\partial\hat u}{\partial x_p}\right|^2\right)dx,
\end{equation}
we get from  (\ref{eq:4.52}),  (\ref{eq:4.53}),  (\ref{eq:4.54}):
\begin{equation}                               \label{eq:4.55}
\left\|\hat x_n^{\frac{1}{2}}\frac{\partial^2(\chi\hat u)}{\partial\hat x_n\partial  x_p}
\right\|_0^2
+\sum_{j=1}^{n-1}
\left\|\frac{\partial^2(\chi\hat u)}{\partial x_p\partial x_j}\right\|_0^2
\leq C_\e+
\e\left(\sum_{j=1}^{n-1}\left\|\frac{\partial^2(\chi \hat u)}{\partial x_j\partial x_p}\right\|_0^2
+\left\|\hat x_n\frac{\partial^2(\chi \hat u)}{\partial\hat x_n\partial  x_p}\right\|^2\right).
\end{equation}
Therefore
\begin{equation}                               \label{eq:4.56}
\left\|\hat x_n^{\frac{1}{2}}\frac{\partial^2(\chi\hat u)}{\partial\hat x_n\partial  x_p}
\right\|_0^2
+\sum_{j=1}^{n-1}
\left\|\frac{\partial^2(\chi\hat u)}{\partial x_p\partial x_j}\right\|_0^2
\leq C_\e,\ \ \ 1\leq p\leq n-1.
\end{equation}
This is true for any $\chi(x)\in C_0^\infty(U)$  in local system of
coordinates,  where $U$  is a neighborhood of any point $x\in S_0$.  
We want to prove that $u\in H^2(\Omega_\delta)$.  It is enough to prove that 
$\frac{\partial^2\hat u}{\partial \hat x_n^2}\in L^2$  since this 
together with (\ref{eq:4.56})
implies that
$\frac{\partial^2\hat u}{\partial\hat x_n\partial x_j}\in L^2,\ 1\leq j\leq n-1$.
Moreover,  for any $\e>0$  we have
\begin{equation}                                       \label{eq:4.57}
\left\|\frac{\partial^2\hat u}{\partial\hat x_n\partial x_j}\right\|_0^2
\leq \e\left\|\frac{\partial^2\hat u}{\partial \hat x_n^2}\right\|_0^2 + 
C_\e\left\|\frac{\partial^2\hat u}{\partial x_j^2}        \right\|_0^2.
\end{equation}
To prove that$\frac{\partial^2 \hat u}{\partial \hat x_n^2}\in L^2$  
we shall again use an estimate of the form
(\ref{eq:4.47})  with $u$ replaced by $Hu$  (cf. [DR]).  Commuting $\qed_g u=0$  and $H$
we get
\begin{equation}                              \label{eq:4.58}
\qed_g Hu+L_2u=0,
\end{equation}
where 
$L_2=H\qed_g-\qed_gH.$                       

Replacing $u$  by $Hu$  we get,  instead of (\ref{eq:4.27}):
\begin{eqnarray}                                  \label{eq:4.59}
0=\frac{1}{2}\int_{\Omega_{ext}}(g^{00}Hu_{x_0}+H^2u)^2+e_{1x_0}(Hu))\sqrt{|g|}\ dx{\Big|}_0^T 
\\
\nonumber
+B(Hu)+T(Hu)
+\int_0^T(L_2u,g^{00}Hu_{x_0}+H^2u)dx_0.
\end{eqnarray}
Note that  the norm
\begin{equation}                                    \label{eq:4.60}
\int_{\Omega_{\delta}}\hat e_{1x_0}(\hat H \hat u)\sqrt{|\hat g|}\ d\hat x<+\infty
\end{equation}
is equivalent  to the norm in $H^2(\Omega_{\delta})$  since  (\ref{eq:4.60})  implies that
$\frac{\partial}{\partial \hat x_j}\hat H\hat u\in L^2,\ 1\leq j\leq n,$  and therefore,
using (\ref{eq:4.56}),  we get that 
$\frac{\partial^2 \hat u}{\partial \hat x_j\partial \hat x_n}\in L^2,\ 1\leq j\leq n$.
We have
\begin{equation}                                 \label{eq:4.61}
\sum_{j,k=1}^n\hat u_{\hat x_j\hat x_k}^2
+C\hat e_{1x_0}(\hat u)\geq \hat e_{1x_0}(\hat H\hat u).
\end{equation}
Analogously to (\ref{eq:4.40})  we have,  using (\ref{eq:4.61}):
\begin{eqnarray}                                \label{eq:4.62}
\hat u_{\hat x_n^2}^2=
\hat u_{\hat x_n^2}^2+ (g^{00}\hat H u_{x_0}+\hat H^2 \hat u)^2 -(g^{00}\hat H\hat u_{x_0}+\hat H^2
\hat u)^2
\\
\nonumber
\geq
C\hat e_{2x_0}(\hat u)-C_1 \hat e_{1x_0}(\hat u)-C_2\hat e_{1x_0}(\hat u_{x_0})
-C_3\sum_{j=1}^{n-1}\sum_{k=1}^n\hat u_{x_j\hat x_k}^2,
\end{eqnarray}
where
\begin{equation}                                 \label{eq:4.63}
\hat e_{2x_0}(\hat u)=(g^{00}\hat H\hat u_{x_0}+ \hat H^2 \hat u)^2+\hat e_{1x_0}(\hat H \hat u).
\end{equation}
Denote
\begin{equation}                                  \label{eq:4.64}
E_2(x_0)=\frac{1}{2}\int_{\Omega_{ext}}e_{2x_0}(u)\sqrt{|g|}\ dx.
\end{equation}
Analogously to (\ref{eq:4.35})  we shall prove that 
\begin{equation}                                 \label{eq:4.65}
T_6(Hu)\geq c\int_0^T E_{2x_0}(u)dx_0-CT.
\end{equation}
As in (\ref{eq:4.39}),  using that 
$(\hat H\hat u)_{\hat x_n}=\hat g^{n0}\hat u_{\hat x_n^2}^2+...$,
  we have:
\begin{eqnarray}                               \label{eq:4.66}
\sum_{j,k=1}^n\hat H(\hat g^{jk})(\hat H\hat u)_{x_j}(\hat H\hat u)_{x_k}
\\
\nonumber
\geq c \hat u_{\hat x_n^2}^2-C_1\sum_{j=1}^{n-1}\hat u_{\hat x_n x_j}^2
-C_2\sum_{j,k=1}^{n-1}\hat u_{x_jx_k}^2- C_3\hat e_{1x_0}(\hat u).
\end{eqnarray}
Since (\ref{eq:4.62})  and (\ref{eq:4.66})  hold in a neighborhood of any point $x\in S_0$
  we get,  using (\ref{eq:4.56})  and (\ref{eq:4.57}),  that
\begin{equation}                                 \label{eq:4.67}
\int_{\Omega_\delta}\sum_{j,k=1}^nH(g^{jk})(Hu)_{x_j}(Hu)_{x_k}\sqrt{|g|}\ dx
\geq  c\int_{\Omega_\delta}e_{2x_0}(Hu)\sqrt{|g|}\ dx - C.
\end{equation}
It follows from (\ref{eq:4.12})  that $u\in H^2(\Omega_{\delta_1}^{(1)})$  for any
$\delta_1>0$  and $\|u\|_{2,\Omega_{\delta_1}^{(1)}}\leq C_{\delta_1}$  Therefore
\begin{equation}                             \label{eq:4.68}
\int_{\Omega_{\delta_1}^{(1)}}\sum_{j,k=1}^nH(g^{jk})(Hu)_{x_j}(Hu)_{x_k}\sqrt{|g|}\ dx
\geq - C\|u\|_{2,\Omega_{\delta_1}^{(1)}} \geq - C C_{\delta_1}.
\end{equation}
Adding (\ref{eq:4.67}) and  (\ref{eq:4.68})  and integrating in $x_0$  we get
(\ref{eq:4.65}).

Note that there is no term in $\sum_{k=1}^5T_k(Hu)$  having the form $c(x)\hat u_{\hat x_n^2}^2$  in
local coordinates.  Any term in $\sum_{k=1}^5 T_k(Hu)$  contains at least  one derivative
in $x_j,\ 1\leq j\leq n-1$.  Therefore, applying inequality (\ref{eq:4.57}),  we get:
\begin{equation}                                       \label{eq:4.69}
\sum_{k=1}^5|T_k(Hu)|\leq \e\int_0^T E_{2x_0}(x_0)dx_0+C_\e T.
\end{equation}
Consider now $\int_0^T(L_2u,g^{00}Hu_{x_0}+H^2u)dx_0$.
This integral contains exactly one term of the form $c(x)\hat u_{\hat x_n^2}^2$   and it has the form:                                                       
\begin{equation}                                 \label{eq:4.70}
(\hat H\hat g^{nn})\hat u_{\hat x_n^2}\hat  H^2\hat u=
(\hat H\hat g^{nn})(\hat g^{n0})^2\hat u_{\hat x_n^2}^2+...
\end{equation}
Since $\hat H\hat g^{nn}>0$  (c.f.  (\ref{eq:4.38}))  we have exactly  the same situation as for 
$T_6(Hu)$.  Therefore
\begin{equation}                                 \label{eq:4.71}
\int_0^T(L_2u,g^{00}Hu_{x_0}+H^2u)dx_0\geq 
c\int_0^TE_{2x_0}(x_0)dx_0-C T.
\end{equation}
Applying (\ref{eq:4.59})  to the interval $(\tau,t)$  and using 
(\ref{eq:4.65}),  (\ref{eq:4.69})  and (\ref{eq:4.71})  we get
\begin{equation}                                   \label{eq:4.72}
E_2(t)-E_2(\tau)+B(Hu)+2c\int_\tau^t E_2(x_0)dx_0\leq CT.
\end{equation}
Therefore, as in (\ref{eq:4.47}),
we have
$$
E_2(t)\leq E_2(0)+C_1,
$$
i.e.  we proved that $u\in H^2(\Omega_{ext})$  and $\|u\|_2\leq C$.

By the Sobolev's embedding theorem
\begin{equation}                         \label{eq:4.73}
|u(x,t)|\leq C\ \ \ \mbox{when}\ \ n=3.
\end{equation}
One can repeat the above arguments to show that $u\in H^m(\Omega_{ext}),m>2$  under the
assumption that the initial data are smooth enough.

\end{document}